# DYNAMIC ENTERPRISE ARCHITECTURE CAPABILITIES AND ORGANIZATIONAL BENEFITS: AN EMPIRICAL MEDIATION STUDY

*Research paper*

Van de Wetering, Rogier, Open University, the Netherlands, rogier.vandewetering@ou.nl

## Abstract

*In recent years the literature has put a greater emphasis on theory building in the context of Enterprise Architecture (EA) research. Specifically, scholars tend to focus on EA-based capabilities that organize and deploy organization-specific resources to align strategic objectives with the particular use of technology. Despite the growth in EA studies, substantial gaps remain in the literature. The most noteworthy gaps are that the conceptualization of EA-based capabilities still lacks a firm base in theory and that there is no conclusive evidence on how EA-based capabilities drive business transformation and deliver benefits to the firm. Therefore, this study focusses on EA-based capabilities, using the dynamic capabilities view as a theoretical foundation, develops and tests a new research model that explains how dynamic enterprise architecture capabilities lead to organizational benefits. Hypotheses associated with the research model are tested using a dataset that contains responses from 299 CIO's, IT managers, and lead architects. Results show that dynamic enterprise architecture capabilities positively influence firms' process innovation and business-IT alignment. These mediating forces are both positively associated with organizational benefits. This study advances our understanding of how to efficaciously delineate dynamic enterprise architecture capabilities in delivering benefits to the organization.*

## 1 Introduction

Global technology trends like big data, the Internet of Things, and the rise of artificial intelligence are making firms' ability to change and adapt its organizations' structure, architecture, and people as crucial as its competitive strategy. These external forces and technology disruptors enact massive transformation changes within firms' business ecosystems, business units, and functions and provide an opportunity to build capabilities in parallel with implementing a new strategic direction. Hence, firms need to accelerate the development of adaptive capabilities to ensure the business can meet the needs of an increasingly complex environment. Moreover, firms need to embrace the business transformation journey to become a top performer in the digital economy, i.e., to become "future-ready" (Weill & Woerner, 2018).

The increased frequency and speed of business-driven and information technology (IT)-driven change opportunities stress the importance of close alignment of IT resources, assets, and capabilities with business processes (Hazen, Bradley, Bell, In, & Byrd, 2017; Ross, Weill, & Robertson, 2006; Shanks, Gloet, Someh, Frampton, & Tamm, 2018). Enterprise Architecture (EA) can be considered a representation of a high-level view of enterprise business processes and the firms' IT systems, their interrelationships, and the extent to which different parts share these systems and processes within the enterprise (Tamm, Seddon, Shanks, & Reynolds, 2011). Within firms, EA is used to provide firms with value across departments, processes, and business units in the process of aligning strategic business objectives with state-of-the-art technologies (Gong & Janssen, 2019; Hazen et al., 2017; Ross et al., 2006; Tamm et al., 2011). Despite the richness of EA studies over the past three decades, there is an inadequate understanding of the value created by EA and so-called EA-based capabilities, that organ-ize and deploy organization-specific resources to align strategic objectives with the particular use of technology (Kotusev, 2019; Lange, Mendling, & Recker, 2016; Shanks et al., 2018; Tamm et al., 2011; Vessey & Ward, 2013). Therefore, there is no conclusive evidence on how EA-based capabilities enable business transformations and deliver organizational benefits in constantly-changing business environments (Hinkelmann et al., 2016; Pattij, Van de Wetering, & Kusters, 2019; Vessey & Ward, 2013).

This study follows previous EA-based capability scholarship that used the dynamic capabilities view (DCV) (Abraham, Aier, & Winter, 2012; Hazen et al., 2017; Labusch, Aier, & Winter, 2013; Shanks et al., 2018; Toppenberg, Henningsson, & Shanks, 2015). The DCV provides a strong theoretical foundation and is accompanied by empirically-validated constructs and items (Hazen et al., 2017; Shanks et al., 2018; Van de Wetering, 2019a). Hence, this study contends that dynamic enterprise architecture capabilities, conceptualized as dynamic capabilities (Teece, 2007; Van de Wetering, 2019a; Van de Wetering, 2019b), allow firms to sense business and IT opportunities and threats (Shanks et al., 2018; Toppenberg et al., 2015).

Moreover, these capabilities allow firms to respond to these opportunities in the ever-changing economic environment (Overby, Bharadwaj, & Sambamurthy, 2006; Sambamurthy, Bharadwaj, & Grover, 2003; Shanks et al., 2018). The literature argues that this particular ability to cultivate the EA to reconfigure the business successfully and the IS/IT landscape, and to adjust for and respond to unexpected changes is an essential driver for 'business-IT alignment' (Gregor, Hart, & Martin, 2007; Korhonen & Molnar, 2014). Also, this study argues that dynamic enterprise architecture capabilities enable firms to gain access to previously unavailable EA resources and sets of decision options which can ultimately enhance their ability to innovate using EA and contribute to organizational benefits (Drnevich & Kriauciunas, 2011; Eisenhardt & Martin, 2000).

The extant literature proclaims that process innovation, i.e., the process view of the business with the application of innovation to the firm's business processes, requires firms to (re)deploy IS/IT and other technologies to enhance the efficiency of new product development and commercialization (Das & Joshi, 2012). Hence, this study and tries to unfold the effect of dynamic enterprise architecture capabilities on organizational benefits. It is hypothesized that this particular effect is indirect and is, therefore, mediated by other intermediate-capabilities and IT-business benefits, i.e., process innovation and business-IT alignment. Therefore, motivated by the call to provide evidence on how EA-based capabilities drive business transformation and deliver benefits, this study tries to answer two closely related research questions:

*(1) Do dynamic enterprise architecture capabilities lead to organizational benefits? And*

*(2) Through which mechanisms of mediating forces are organizational benefits achieved?*

This study proceeds as follows. First, this study outlines the theoretical ground and hypotheses development. Then, this work outlines the study design and methods. The final sections present the results and discuss the essential outcomes of this study. This study ends with some concluding remarks.

## 2 Theoretical ground and model development

This study builds upon the foundation of the resource-based view of the firm (RBV) and the DCV as its theoretical foundation. Specifically, the DCV is embraced as a theory to examine the proposed impact of dynamic enterprise architecture capabilities on organizational benefits.

The RBV, as a resource-synchronization theory, tries to explain how a firm can build and deploy its unique resources (tangible and non-tangible) to obtain a competitive position and business value (Tarafdar & Gordon, 2007). Within the context of EA, the RBV argues that investments in the organization's EA artifacts (e.g., content, standards, models) and services are crucial in the process of developing and aligning organizational capabilities (Mithas, Tafti, Bardhan, & Goh, 2012; Schelp & Stutz, 2007; Shanks et al., 2018; Sheikh, Sood, & Bates, 2015; Someh, Frampton, Davern, & Shanks, 2016). Drawing from the RBV logic, various IS studies have postulated that the targeted use of EA assets and resources can provide firms with a competitive edge (Someh et al., 2016; Wade & Hulland, 2004).

However, competitive benefits and business value are not achieved directly. Instead, the benefits are typically achieved through a series of value-creating activities and transformative capabilities (Shanks et al., 2018; Someh et al., 2016; Wade & Hulland, 2004). Therefore, the RBV acknowledges that the single investment in EA is not a sufficient condition for enhancing organizational benefits. It is thus essential for firms that they identify those organizational capabilities that EA should be infused in to enable business transformations and deliver benefits to the firm (Kim, Shin, Kim, & Lee, 2011; Kohli & Grover, 2008; Van de Wetering, Versendaal, & Walraven, 2018). This is where dynamic enterprise architecture capabilities come into play.

## 2.1 Dynamic Enterprise Architecture Capabilities

The DCV extends the RBV and attempts to explain the processes through which a firm evolves in changing environments and maintains a competitive edge (Schilke, 2014; D. J. Teece, Pisano, & Shuen, 1997). Due to conditions of high environmental uncertainty, market volatility, and frequent change, scholars have raised questions regarding the rate to which traditional operational and existing 'resource-based' capabilities erode and cease to provide competitive gains (Drnevich & Kriauciunas, 2011). Dynamic capabilities are generally considered as the ability of organizations to integrate, reconfigure, gain, and release resources to match and even create market change (Eisenhardt & Martin, 2000; Teece et al., 1997).

In the context of strategic management and IS literature, recently, some researchers argue that EA-based capabilities are valuable to firms in the process of using, deployment, and diffusion of EA in decision-making processes, and the organizational routines that drive IT and business capabilities (Brosius, Aier, Haki, & Winter, 2018; Hazen et al., 2017; Shanks et al., 2018). Moreover, Shanks et al. (2018) argue that EA-based capabilities are essential to leveraging EA advisory services within the firm. Likewise, Hazen et al. (Hazen et al., 2017), following the DCV, provide foundational work that shows that EA-based capabilities can enhance organizational agility and indirectly enhance organizational performance. These outcomes are consistent with work by Foorthuis et al. (Foorthuis, Van Steenbergen, Brinkkemper, & Bruls, 2016) that demonstrate the importance of intermediate EA-enabled outcomes that contribute to the achievement of particular business goals and objectives. Hence, recent EA scholarship argues that complementary EA capabilities enable firms to leverage their EA effectively (Hazen et al., 2017; Tamm et al., 2011), contribute to IT efficiency and IT flexibility (Schmidt & Buxmann, 2011), and can drive alignment between business and IT (Hinkelmann et al., 2016).

This study concurs with this EA-based capability view. It considers dynamic enterprise architecture capabilities as a dynamic capability that helps organizations identify and implement new business and IT initiatives to ensure that the organizations' assets and resources are current with the needs of the business. Following the tenets of the DCV, this study argues that it is likely that the extent to which EAs are leveraged successfully within the organization depends on the dynamic capabilities that collectively use the EA to sense environmental threats and business opportunities, while simultaneously implementing new strategic directions. This study conceives dynamic enterprise architecture capabilities as the firm's ability to exploit its EA to share assets, and recompose and renew organizational resources under rapidly changing internal and external conditions to accomplish strategic objectives and the desired end state ( Van de Wetering, 2019b).

Starting from the conceptualization of dynamic capabilities by (Teece et al., 1997), and subsequently building on previous EA-based capability studies, this study synthesizes the reach and range of dynamic enterprise architecture capabilities through three related, but distinct capabilities, i.e., EA sensing capability, EA mobilizing capability, and EA transformation capability. An EA sensing capability highlights the role of EA in firms' deliberate posture toward sensing and identifying new business opportunities or potential threats and developing a greater reactive and proactive strength in the business domain (Shanks et al., 2018; Toppenberg et al., 2015). An EA mobilizing capability refers to organizations' capability to use EA in the process of evaluating, prioritizing, and selecting potential solutions and mobilize firm resources in line with a potential solution (Overby et al., 2006; Sambamurthy et al., 2003; Shanks et al., 2018). Finally, an EA transforming capability can be considered the ability to use the EA to successfully reconfigure business processes and the technology landscape, to engage in resource recombination and to adjust for and respond to unexpected changes (Drnevich & Kriauciunas, 2011; Mikalef, Pateli, & Van de Wetering, 2016; Pavlou & El Sawy, 2006; Shanks et al., 2018).

## 3 Hypotheses development

### 3.1.1 Dynamic Enterprise Architecture Capabilities and business-IT alignment

Both in scientific literature and practice, it is a well-known fact that achieving a state of business-IT alignment[1] is essential to leverage the maximum potential organizational benefits (Henderson & Venkatraman, 1993; Wu, Straub, & Liang, 2015). Business-IT alignment typically refers to the degree to which the IT strategies, objectives, and priorities appropriately and harmoniously support business strategies, objectives, and priorities (Bradley, Pratt, Byrd, Outlay, & Wynn, 2012; Kearns & Lederer, 2003; Luftman & Kempaiah, 2007). As such, this research focusses on the antecedents and drivers of business-IT alignment and, hence, the content of alignment, i.e., the match between realized business and IT strategy (Wu et al., 2015). This dimension of alignment is classified in the literature as 'intellectual alignment' (Chan & Reich, 2007; Tallon & Pinsonneault, 2011; Wu et al., 2015).

Having a clear overview of the EA resources (e.g., EA content, EA standards, services, and other artifacts), and thus architectural transparency and a planned architectural design, can facilitate the process of integrating IT assets, resources, business processes and services across various architectural layers (Foorthuis et al., 2016; Wang, Zhou, & Jiang, 2008). EA can be leveraged to bridge the communication gap between business and IT stakeholders, facilitate cross-organizational dialogue and input (Tamm et al., 2011), and hence, improve business-IT-alignment (Gregor et al., 2007; Kotusev, 2019). Following the above discussion, it can be argued that dynamic enterprise architecture capabilities allow firms to continuously sense the on-going change within the organizations' internal and external business and IS/IT landscape and adequately respond by mobilizing firm resources to support business processes, specific user needs, and requirements using the EA (Tamm et al., 2011). Hence, this particular ability to cultivate the EA to reconfigure the business successfully and the IS/IT landscape, recombine resource and to adjust for and respond to unexpected changes is an essential driver for business-IT alignment. Hence, as firms proactively invest more in their dynamic enterprise architecture capabilities, one of the results is better business-IT-alignment (Gregor et al., 2007; Korhonen & Molnar, 2014). Hence, the following is defined:

*H1: Dynamic enterprise architecture capabilities have a positive effect on business-IT alignment.*

### 3.1.2 Business-IT alignment and organizational benefits

Achieving a state of alignment comes with many organization benefit gains, including market growth, cost control, financial performance, increasing levels of customer satisfaction and augmented reputation (Bradley et al., 2012; Chan & Reich, 2007; Kearns & Lederer, 2003; Tallon & Pinsonneault, 2011; Wu et al., 2015). Moreover, prior studies suggest that aligning the IT strategy with the business strategy will likely have an impact on process agility and, thus, the way firms can easily and quickly reshape their business processes in turbulent business environments (Tallon, 2007; Tallon & Pinsonneault, 2011).

Although EA facilitates decision-making processes and brings the business and IT investment decisions in closer alignment to the organizational goals (Tamm et al., 2011), EA by itself does not create any value for the firm (Hazen et al., 2017; Shanks et al., 2018). Instead, IT and business managers can

---

[1] This research focusses on measurement of alignment at a single point in time, rather than focusing on a process that evolves over time.

drive enterprise-wide transformational changes and provide the firm with various opportunities to build and deploy and capabilities while also actively practicing the firm's new strategic direction using the EA. Previous EA-based capabilities scholarship shows that many of the benefits of EA are intangible and value is achieved indirectly (Foorthuis et al., 2016; Shanks et al., 2018). This study, therefore, theorizes that the business-IT alignment mediates the relation between dynamic enterprise architecture capabilities and organizational benefits. Business-IT alignment is, thus, a crucial mediating force in the particular chain of EA value creation (Kohli & Grover, 2008) and, therefore, a crucial antecedent of organizational benefits. Therefore, the following hypothesis is defined:

*H2: Business-IT alignment will mediate the relationship between dynamic enterprise architecture capabilities and organizational benefits.*

However, given the unique nature of the dynamic enterprise architecture capabilities—in terms of their reach and range—and their hypothesized relationship with organizational benefits, it is likely that dynamic enterprise architecture capabilities have a positive impact not only on business-IT alignment but also on process innovativeness of the firm. Various scholars argue that process innovation is the outcome of organizational learning and EA resource orchestration for which the roots can be traced back to dynamic capabilities (Breznik & Hisrich, 2014; Giniuniene & Jurksiene, 2015; Niemi & Pekkola, 2017).

### 3.1.3 Dynamic enterprise architecture capabilities and process innovation

There are many other forms of innovation (e.g., business model, leadership) that relate mainly to process innovation (Assink, 2006). This study focuses on process innovation (or 'process innovativeness') as it has a central place in the extant literature, and this type of innovation requires firms to (re)deploy IS/IT and other technologies to enhance the efficiency of new product development and commercialization (Das & Joshi, 2012). Teece et al. (2016) concur with this view, as he notes that strong dynamic capabilities are required for fostering the organizational agility and associated requirements necessary for innovation. To drive process innovation, it is essential that firms re-allocate resources systematically and improve methods in service and production operations through the use of technological advancements (Das & Joshi, 2012).

EA-based sensing capabilities facilitate firms in their process to spot, interpret and pursue new IS/IT and technological innovations (e.g., cloud, IoT, big data analytics, AI, business intelligence), business and process opportunities or identify potential threats (Overby et al., 2006; Pavlou & El Sawy, 2011). They help firms by aligning EA services with key stakeholders' wishes and position EA resources so that targeted efforts for innovation can be initiated. Moreover, EA-based capabilities foster organizational learning by designing both IT and business facets of the enterprise as well as its relationship with the business ecosystems to enable innovation and its ability to adapt in conjunction with the business environment (Korhonen & Molnar, 2014; Lapalme, 2012; Yu, Deng, & Sasmal, 2012).

Once technological and business opportunities are first glimpsed, they must be addressed through maintaining and improving technological competences and complementary firm assets (Pavlou & El Sawy, 2006; Teece, 2007). Hence, an EA-mobilizing capability allows firms to consciously direct investments in adaptiveness of the firm, use EA in the process of evaluating, prioritize and select potential IT and business solutions and mobilize firm resources accordingly (Overby et al., 2006; Sambamurthy et al., 2003; Shanks et al., 2018; Yu et al., 2012). An EA-mobilizing capability is, thus, an essential ingredient for firms that want to adapt its resources and assets to the continually evolving customer wishes, demands and market, and technology trends and shape their environment through innovation (Teece, 2007).

Finally, an EA transforming capability allows firms to engage in recombination and re-deployment of resources, change collaboration within the enterprise and to adjust for and respond to unexpected changes and the need for innovation (Drnevich & Kriauciunas, 2011; Mikalef et al., 2016; Pavlou & El Sawy, 2006; Shanks et al., 2018). Dynamic enterprise architecture capabilities, thus, allow firms to use EA in decision-making processes and support competences to change the position of IS/IT and other firm resources through process innovation (Korhonen & Molnar, 2014; Lapalme, 2012; Roberts,

Galluch, Dinger, & Grover, 2012; Shanks et al., 2018). Hence, by these EA-based capabilities, firms can gain access to previously unavailable EA resources and sets of decision options which can ultimately enhance their ability to innovate using EA and contribute to organizational benefits (Drnevich & Kriauciunas, 2011; Eisenhardt & Martin, 2000). Hence, this study defines:

*H3: Dynamic enterprise architecture capabilities have a positive effect on firms' level of process innovation.*

### 3.1.4 Process innovation and organizational benefits

The level of process innovation tends to rely not on individual resources. Instead, it seems that process innovation is based on unique combinations of complementary resources and cooperation of cohesive units governed by EA-based capabilities (Lange et al., 2016; Schmidt & Buxmann, 2011; Tamm et al., 2011). The literature claims that EA-based capabilities are a precursor for process innovation. Process innovation enabled by dynamic enterprise architecture capabilities, in turn, influences organizational benefits in several ways, as previously documented in the literature (Das & Joshi, 2012; Davenport, 1993). Specifically, process innovation leads to better financial and operational results (e.g., return on investment, market growth, cost reduction) (Hult, Hurley, & Knight, 2004), enhanced levels of productivity, process efficiencies and effectiveness (Gray, Matear, & Matheson, 2002), as well as enhanced levels of customers' perceived value (Das & Joshi, 2012). Hence, the following hypothesis is defined:

*H4: Process innovation will mediate the relationship between dynamic enterprise architecture capabilities and organizational benefits.*

The current research model thus contains four key constructs and the accompanying hypotheses. Figure 1 shows the research model that will be empirically validated.

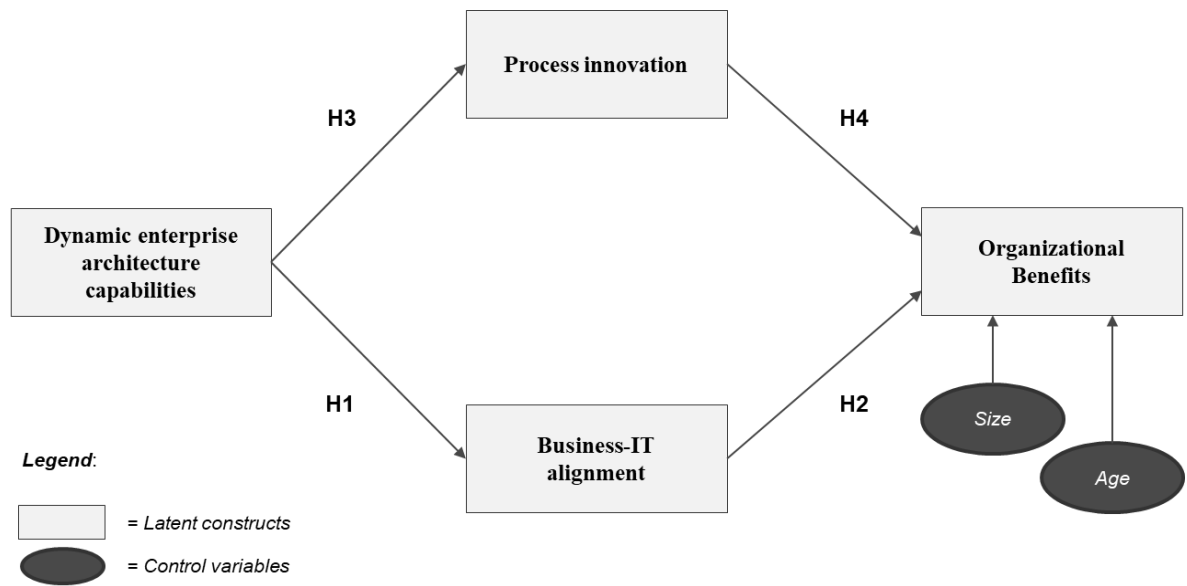

*Figure 1.    Research model*

## 4    Study design and methods

This study embraces a deductive approach that guides the study's design by focussing on the prediction of the key outcome construct in this study. Henceforth, claims are grounded in the existing body of knowledge, and this study also focuses on the development of persuasive arguments to substantiate these claims. As such, the current study required an extensive cross-sectional sample to test the model and the associated hypotheses.

## 4.1     Data collection and sample description

A questionnaire was developed that included 35 main questions covering all relevant constructs in the research model (Tables 1*a-c*). The applied survey was pretested by 13 IS scholars, EA, and IT/business practitioners and Master students to enhance the content and face validity of the survey items. Students of an advanced Business and IT Master course on Enterprise Architecture and organizational capabilities of a Dutch University were asked to participate in this survey. The Netherlands belongs to the top European countries that deliver substantial economic impact using IT. Dutch firms are currently in a very proficient position to make use of the various economic and social opportunities created by digitalization according to the Dutch Digitalisation Strategy[2]. Therefore, they also form a suitable sample and frame of reference for the current investigation. Students were voluntarily asked to fill in the survey from the particular perspective of the organization where they currently work.

These particular Master students are experienced business or IT managers, consultants, and senior practitioners, and therefore, match our respondent profile. They are most likely familiar with the strategic role of EA within the firm. Nonetheless, to ensure a collective and firm-wide view, the respondents were also invited to consult their managers (or any other colleague), in case they were not sure about a particular survey item. Also, all students ($N$=235) had to distribute this survey to two knowledgeable domain experts from other organizations (e.g., CIOs, IT managers, and lead enterprise architects) following a snowball method.

The survey was put to a rigorous pretesting procedure, enhancing both the reliability and validity. Also, construct definitions were provided for the respondents and the survey followed a logical structure. Respondents were offered a research report with the most important outcomes of this study. Anonymity was guaranteed, and respondents could withdraw their scores if they wanted to. The final survey was used to collect data as part of a field study. During the data collection, also various controls were built in so that every organization completed the survey only once. The data collection phase started on the 17th of October 2018 and ended on the 16th of November 2018. A total of 669 unique respondents from different organizations commenced with the survey. After removing cases with either (partly) incomplete ($N$=290) or unreliable values ($N$=80), this study includes a total of 299 usable questionnaires for the analyses. The majority of respondents operate in the private sector (i.e., 57%) and the public sector (i.e., 36%). Only a small percentage (i.e., 7%) comes from other categories such as private-public partnerships, and non-governmental organizations. The dataset can roughly be classified into small to medium-size firms (i.e., 41%, no. of employees < 1000) and large enterprises (i.e., 59%, no. of employees > 1000). The majority of responses come from high to executive managers, i.e., CEOs, CIOs, and IT management (approximately 70%). Approximately 60% of the respondents had more than 11 years of working experience. Of all the respondents, 40% had even more than 20 years of experience.

As this research targets single respondents, there is a possibility that bias might exists. Hence, possible method bias was proactively accounted for to mitigate possible methods effects following guidelines by Podsakoff et al. (2003). Moreover, this study accounts for possible common method variance (CMV) per suggestions of Podsakoff (Podsakoff et al., 2003). T-tests group analyses for early (first two weeks) and later responses (final two weeks) for each research construct showed no significant differences showing that possible non-response bias is not present. Finally, Harman's single factor test was performed using IBM SPSS Statistics™ v24 on the study constructs. Hence, the construct variables were all loaded on to a single construct in an Exploratory Factor Analysis (EFA). Outcomes of this analysis showed that no single factor attributes to the majority of the variance; the sample is not affected by CMB (Podsakoff et al., 2003).

---

[2] This report, as retrieved from https://www.government.nl/, is developed by Ministry of Economic Affairs and Climate Policy of the Netherlands. The report reflects on what is needed for the Netherlands to be ready for the digital future.

## 4.2 Constructs and measurement items

This study attempted to include existing validated measures where possible. First, this study newly conceptualizes the EA-based capability as a dynamic capability. Hence, this study adopts three elementary routines for dynamic capabilities, i.e.: (1) EA-sensing capability, (2) EA-mobilizing capability, (3) EA-transforming capability (Van de Wetering, 2019a; Van de Wetering, 2019b). These underlying capabilities collectively form the dynamic enterprise architecture capabilities construct. A rigorous conceptual and theoretical development should precede the development of this construct. This research followed a staged approach in the development of this new multi-item scale following established guidelines (MacKenzie, Podsakoff, & Podsakoff, 2011). First, measurement items were directed from either previously cited in or implied by extant conceptual and empirical work (Drnevich & Kriauciunas, 2011; Mikalef et al., 2016; Overby et al., 2006; Pavlou & El Sawy, 2011; Sambamurthy et al., 2003; Shanks et al., 2018; Wilden, Gudergan, Nielsen, & Lings, 2013). The first pool of scale items was developed using a seven-point Likert-type scale, ranging from ''strongly disagree'' to ''strongly agree.'' Then, two sub-phases of scale development and purification followed based on previously outlined recommendations (MacKenzie et al., 2011), i.e., item-sorting analysis and expert reviews. The item-to-construct sorting approach was employed to establish tentative item reliability and validity (Moore & Benbasat, 1991), while expert reviews once more evaluated all the established scale items and offered improvement suggestions (Presser et al., 2004). These two sub-phases enhanced the reliability and construct validity of dynamic enterprise architecture capabilities at a pre-testing stage. The results of these intensive phases are omitted for the sake of brevity. This second-order construct is modeled using the reflective-formative type II model (Becker, Klein, & Wetzels, 2012; Jarvis, MacKenzie, & Podsakoff, 2003).

| Construct | | Measurement item | $\lambda$ | $\mu$ | *Std.* |
|---|---|---|---|---|---|
| | *To what extent do you agree with the following statements? (1 – strongly disagree 7 – strongly agree). Mobilizing and transforming capability use the same Likert Scale.* | | | | |
| Sensing capability | S1 | We use our EA to identify new business opportunities or potential threats | 0.77 | 3.83 | 1.61 |
| | S2 | We review our EA services regularly to ensure that they are in line with key stakeholders wishes | 0.84 | 4.10 | 1.60 |
| | S3 | We adequately evaluate the effect of changes in the baseline and target EA on the organization | 0.86 | 4.02 | 1.48 |
| | S4 | We devote sufficient time to enhance our EA to improve business processes | 0.82 | 4.01 | 1.56 |
| | S5 | We develop greater reactive and proactive strength in the business domain using our EA | 0.85 | 4.04 | 1.54 |
| Mobilizing capability | M1 | We use our EA to draft potential solutions when we sense business opportunities or potential threats | 0.85 | 4.39 | 1.51 |
| | M2 | We use our EA to evaluate, prioritize and select potential solutions when we sense business opportunities or potential threats | 0.86 | 4.37 | 1.51 |
| | M3 | We use our EA to mobilize resources in line with a potential solution when we sense business opportunities or potential threats | 0.88 | 4.19 | 1.45 |
| | M4 | We use our EA to draw up a detailed plan to carry out a potential solution when we sense business opportunities or potential threats | 0.87 | 4.12 | 1.59 |
| | M5 | We use our EA to review and update our practices in line with renowned business and IT best practices when we sense business opportunities or potential threats | 0.84 | 4.22 | 1.48 |
| Trans. capability | T1 | Our EA enables us to successfully reconfigure business processes and the technology landscape to come up with new or more productive assets | 0.85 | 4.40 | 1.45 |
| | T2 | We successfully use our EA to adjust our business processes and the technology landscape in response to competitive strategic moves or market opportunities | 0.87 | 4.17 | 1.56 |
| | T3 | We successfully use our EA to engage in resource recombination to match our product-market areas and our assets better | 0.83 | 3.95 | 1.47 |
| | T4 | Our EA enables flexible adaptation of human resources, processes, or the technology landscape that leads to competitive advantage | 0.84 | 3.88 | 1.50 |
| | T5 | We successfully use our EA to create new or substantially changed ways of achieving our targets and objectives | 0.87 | 4.06 | 1.51 |
| | T6 | Our EA facilitates us to adjust for and respond to unexpected changes | 0.80 | 4.02 | 1.46 |

*Table 1a.        Constructs and measurement items for dynamic enterprise architecture capabilities*

*Business/IT-alignment* is measured as a reflective first-order construct following (Bradley et al., 2012; Chan, 2002) containing three existing items to capture the firms' IT strategic alignment between the business and IT plans, priorities, and strategies. *Process innovation* is likewise modeled as a reflective first-order construct following (Prajogo & Sohal, 2003). Relevant aspects include the extent to which firms have technological competitiveness and the novelty of technology used in critical processes.

| Construct | | Measurement item | λ | μ | Std. |
|---|---|---|---|---|---|
| Alignment | | *Please choose the appropriate response for each item (1 – strongly disagree 7 – strongly agree)* | | | |
| | A1 | Our organization has a business plan to use existing technology to enter new market segments | 0.81 | 4.31 | 1.63 |
| | A2 | Our organization has a business plan to develop new technologies for new kinds of products/services | 0.81 | 4.61 | 1.61 |
| | A3 | Business and IT strategies are consistent | 0.81 | 4.41 | 1.52 |
| Process inn. | | *How would you rate your organization's process innovation capabilities in comparison to the main competitors in the same industry? (1 = much weaker than competition; 7 = much stronger than competition)?* | | | |
| | P1 | The technological competitiveness | 0.84 | 4.67 | 1.33 |
| | P2 | The updated-ness or novelty of technology used in key processes | 0.88 | 4.55 | 1.31 |
| | P3 | The speed of adoption of the latest technological innovations in key processes | 0.88 | 4.26 | 1.42 |
| | P4 | The rate of change in key processes, techniques, and technology | 0.88 | 4.19 | 1.36 |

Table 1b.   Constructs and measurement items for the mediating forces, i.e., business/IT-alignment, and process innovation.

This study follows Shanks et al. (2018) for the multi-dimensional nature of organizational benefits. Hence, the current study considers organizational benefits to be the long term firm benefits that result from intermediate-capabilities and IT-business benefits. Organizational benefits are conceptualized as a second-order factor using the reflective-formative type II model and contains three underlying first-order benefits factors, i.e., process agility (Tallon & Pinsonneault, 2011), competitive advantage (Chen et al., 2014; Rai & Tang, 2010) and increased value (Chen & Tsou, 2012). Process agility concerns the firms' "ability to detect and respond to opportunities and threats with ease, speed, and dexterity" (Tallon & Pinsonneault, 2011, p. 464). This study used five validated items from Tallon and Pinsonneault (2011). Competitive advantage has several dimensions, including a higher return on investment than competitors, better growth of market share than competitors, and better profitability. Finally, increased value is measured through customer satisfaction, customer loyalty and business brand and image in comparison to competitors. This study controlled for possible confounding relationships by adding several widely-used control variables in IS research, i.e., firm size and age.

| Construct | | Measurement item | λ | μ | Std. |
|---|---|---|---|---|---|
| Organizational benefits | | *How would you rate your firm's process agility aspects in comparison to industry competitors? (1. Much weaker than the competition–7. Much stronger than the competition)?* | | | |
| | AG1 | Expanding into new regional or international markets | 0.70 | 4.35 | 1.33 |
| | AG2 | Responsiveness to customers | 0.81 | 4.71 | 1.22 |
| | AG3 | Responsiveness to changes in market demand | 0.88 | 4.55 | 1.17 |
| | AG4 | Customization of products or services to suit indiv. customers | 0.68 | 4.87 | 1.28 |
| | AG5 | Adopt new technologies to produce better, faster and cheaper products and services | 0.70 | 4.40 | 1.30 |
| | | *Please choose the appropriate response for each item (1 – strongly disagree 7 – strongly agree). During the last 2 or 3 years we perform much better than our main competitors in the same industry in:* | | | |
| | CA1 | Growth in market share | 0.86 | 4.65 | 1.33 |
| | CA2 | Profitability | 0.91 | 4.54 | 1.35 |
| | CA3 | Sales growth | 0.91 | 4.54 | 1.33 |
| | CA4 | Return on investment (ROI) | 0.84 | 4.41 | 1.29 |
| | VL1 | Increasing customer satisfaction | 0.91 | 4.88 | 1.27 |
| | VL2 | Increasing customer loyalty | 0.92 | 4.76 | 1.27 |
| | VL3 | Enhancing business brand and image | 0.87 | 4.84 | 1.34 |

Table 1c.   Constructs and measurement items for organizational benefits.

*Van de Wetering / Dynamic enterprise architecture capabilities and benefits*Tables 1*a-c* show all included measurement items, their respective item-to-construct loadings (λ), mean values (μ), and also the standard deviations (std.). Each item in the final survey was measured using a seven-point Likert scale (1: strongly disagree to 7: strongly agree).

## 4.3  Model estimations

This study relied on the use of SmartPLS version 3.2.7. (Ringle, Wende, & Becker, 2015), which is a Structural Equation Modeling (SEM) application using Partial Least Squares (PLS) to estimate the research model and run parameter estimates. PLS-SEM is variance-based and is considered a better alternative than covariance-based modeling techniques (e.g., LISREL, AMOS) when the emphasis is on prediction since PLS tries to maximize the explained variance in the dependent construct (Chin, 1998; Hair Jr, Hult, Ringle, & Sarstedt, 2016). Also, PLS readily handles both reflective and formative measures (Ringle, Sarstedt, & Straub, 2012) (Hair Jr, Sarstedt, Ringle, & Gudergan, 2017) as is the case in this research, and PLS provides researchers with a greater ability to predict and understand the role and formation of latent constructs and their relationships among each other (Chin, 1998; Hair Jr et al., 2016; Ringle et al., 2012).

Analyses make use of the path weighing scheme within SmartPLS. Also, a non-parametric bootstrapping procedure was employed to compute the level of the significance of the regression coefficients running from the first-order constructs to the second-order construct. In this process, 5000 replications were used to obtain stable results and to interpret their significance. Finally, the 299 organizations in the dataset far exceed all minimum requirements to run the SEM analyses (Hair, Ringle, & Sarstedt, 2011; Hair Jr et al., 2017).

# 5  Empirical validation

## 5.1  Evaluation of the outer model

This study subjects the research model's constructs to internal consistency reliability, convergent validity, and discriminant validity test through SmartPLS (Ringle et al., 2015). At the construct level, a composite reliability (CR) assessment was employed (Hair Jr et al., 2016). Typically, CR values should be above 0.70, as is the case in this research (see Table 2). Also, this study assessed construct-to-item loadings. None of the items had to be removed as all loadings were above 0.70 (Fornell & Bookstein, 1982)[3]. Next, this study assessed convergent and discriminant validity (Hair Jr et al., 2016; Hair Jr et al., 2017). Hence, convergent validity was assessed by examining if the average variance extracted (AVE) is above the generally accepted lower limit of 0.50 (Fornell & Larcker, 1981). All the obtained AVE values exceed the minimum threshold value. In a subsequent step, this study assessed the discriminant validity through three different, but related tests. First, the data were assessed to detect high-loadings on the hypothesized constructs and low cross-loadings (i.e., correlations) on other constructs (Farrell, 2010). The data showed that all items load more strongly on their intended latent constructs than they correlate on other constructs. Second, the Fornell-Larcker criterion was assessed. In doing so, PLS was used to investigate if the square root of the AVEs of all constructs was larger than the cross-correlation (see the diagonal entries in bold in Table 2).

All square root values are higher than the shared variances of the constructs with other constructs in the model (Hair Jr et al., 2017). Finally, this study found additional evidence for discriminant validity by employing the relatively newly developed heterotrait-monotrait (HTMT) metric (Henseler, Ringle, & Sarstedt, 2015). All values showed acceptable outcomes far below the conservative 0.90 upper bound. As shown in Table 2, the first-order reflective measures are valid and reliable. All first-order constructs demonstrate a significant relationship with their respective higher-order construct (i.e., dy-

---

[3] Only one measurement item in the survey had a loading of 0.68; this is still in the range of acceptable item loadings.

*Twenty-Eigth European Conference on Information Systems (ECIS2020), Marrakesh, Morocco.*       10

namic enterprise architecture capabilities and organizational benefits). Also, the assessed variance inflation factors (VIFs) are well below a conservative critical value of 3.5. These outcomes, in addition to the absence of non-significant relations between first-order capabilities and the second-order constructs, indicate that no multicollinearity exists within our model (Kock & Lynn, 2012).

|      | EAS   | EAM   | EAT   | BIA   | PI    | VL    | CA    | PA    |
|------|-------|-------|-------|-------|-------|-------|-------|-------|
| EAS  | **0.829** |       |       |       |       |       |       |       |
| EAM  | 0.782 | **0.857** |       |       |       |       |       |       |
| EAT  | 0.784 | 0.780 | **0.843** |       |       |       |       |       |
| BIA  | 0.402 | 0.436 | 0.454 | **0.808** |       |       |       |       |
| PI   | 0.253 | 0.247 | 0.407 | 0.415 | **0.872** |       |       |       |
| VL   | 0.215 | 0.214 | 0.264 | 0.372 | 0.330 | **0.900** |       |       |
| CA   | 0.214 | 0.275 | 0.221 | 0.374 | 0.330 | 0.628 | **0.875** |       |
| PA   | 0.235 | 0.258 | 0.324 | 0.427 | 0.576 | 0.486 | 0.433 | **0.828** |
|      |       |       |       |       |       |       |       |       |
| AVE  | 0.686 | 0.734 | 0.710 | 0.652 | 0.760 | 0.811 | 0.765 | 0.686 |
| CR   | 0.916 | 0.932 | 0.936 | 0.849 | 0.927 | 0.928 | 0.929 | 0.897 |

*Note: EAS—EA sensing capability; EAM—EA mobilizing capability; EAT—EA transforming capability; BIA—Business-IT alignment; PI—Process innovation; VL—Value; CA—Competitive advantage; PA—Process agility.*

*Table 2.        Convergent and discriminant validity assessment of all the first-order reflective constructs*

The research models' fit, predictive relevance, and the structural model can now be evaluated.

## 5.2  Evaluation of the inner model and hypotheses testing

The literature proposes the Standardized Root Mean Square Residual (SRMR) as a new model fit index. It calculates the difference between observed correlations and the model's implied correlations matrix (Hair Jr et al., 2016; Hu & Bentler, 1999). Hence, this study checks the model by assessing the model fit before further assessing the structural model and associated hypotheses. Current model fit indices should, however, be interpreted with caution as these metrics are not fully established PLS-SEM evaluation criteria. The obtained 0.069 is below the conservative 0.08 mark that is proposed by (Hu & Bentler, 1999). As a final step, the model's predictive relevance is calculated using the $Q^2$ of our endogenous constructs (i.e., using Stone–Geisser's test). All $Q^2$ values are above the threshold value of zero, thereby indicating the overall model's predictive relevance. The structural model and the hypothesized relationships among the model's constructs can now be assessed.

The structural model explains 29% of the variance for organizational benefits ($R^2$=.29), after removing all non-significant relationships from the model. This outcome is considered a moderate effect (Chin, 1998). Also, dynamic enterprise architecture capabilities explain 22.0% of the variance in Busines-IT alignment ($R^2$=.22) and 12% (i.e., $R^2$=.12) of the variance in process innovation. Overall, these coefficients of determination support the explanatory power of the research model, next to the model fit indices and the obtained significant path coefficients ($p < 0.0001$). Table 3 summarizes the structural model assessment findings and additionally shows the estimated effect sizes[4] ($f^2$), and the confidence intervals (Lower bound, 0.5%, – Upper bound, 99.5%) of the structural model analyses.

---

[4] With effect sizes, the specific contribution of particular exogenous constructs to an endogenous latent constructs $R^2$ can be determined.

### 5.2.1 Mediation analyses

This study followed specific guidelines by (Baron & Kenny, 1986; Hair Jr et al., 2016; Hayes, 2013) for multiple mediation analysis procedures to address the imposed mediation effects within the research model specifically. First, the direct effect of dynamic enterprise architecture capabilities on organizational benefits is both positive and significant ($β=0.31$, $t=5.398$, $p ≤ 0.0001$). Hence, this fulfills the first mediation condition, as suggested by Kenny (Baron & Kenny, 1986). Next, the significance of the indirect effects (i.e., mediating paths) was integrally established (i.e., simultaneous consideration of all mediating constructs) through a bootstrapping approach using a non-parametric resampling procedure (Hair Jr et al., 2016; Hayes, 2013). Then, the included the direct path (dynamic enterprise architecture capabilities → Organizational benefits) showed a non-significant relationship ($β=0.06$, $t=1.112$, $p = 0.26$). The specific indirect effects (DEAC → BIA → OB and DEAC → PI → OB, see Table 3) should be interpreted as the indirect effect of DEAC on OB through a given mediation construct (i.e., BIA or PI) while controlling for the other mediating constructs. This study concludes that full mediation characterizes this current structural model. This study, therefore, finds support for the four hypotheses, while all included control variables showed non-significant (n.s.) effects (see Table 3).

| Model path | Effect | Effect size ($f^2$) | Confidence interval (bias corrected) | $t$-value | Sign. | Conclusion |
|---|---|---|---|---|---|---|
| DEAC → BIA | 0.468 | 0.280 | CI (0.329 – 0.579) | 9.791 | YES | H1 Supported |
| BIA → OB | 0.312 | 0.114 | CI (0.110 – 0.454) | 5.305 | YES | |
| DEAC → BIA → OB (*mediation by BIA*) | 0.146 | - | CI (0.068 – 0.229) | 4.688 | YES | H2 Supported |
| DEAC → PI | 0.335 | 0.128 | CI (0.196 – 0.451) | 6.620 | YES | H3 Supported |
| PI → OB | 0.332 | 0.130 | CI (0.185 – 0.467) | 5.848 | YES | |
| DEAC → PI → OB (*mediation by PI*) | 0.111 | - | CI (0.054 – 0.187) | 4.284 | YES | H4 Supported |
| DEAC → OB | 0.063 | 0.004 | CI (-0.074 – 0.2147) | 1.117 | NO | No direct effect |
| Size → OB | 0.028 | 0.002 | CI (-0.066 – 0.158) | 0.521 | NO | No confounding |
| Age → OB | -0.065 | -0.005 | CI (-0.184 – 0.039) | 1.207 | NO | No confounding |

*Note: DEAC—Dynamic enterprise architecture capabilities; OB—Organizational benefits; BIA—Business-IT alignment; PI—Process innovation*

*Table 3.     Summary of the hypotheses testing and structural model outcomes*

## 6 Discussion and conclusion

### 6.1 Theoretical and practical contributions

Motivated by the call to provide empirical evidence on how EA-based capabilities drive business transformation and deliver benefits, this study shows how dynamic enterprise architecture capabilities bring benefits to the firm using data from 299 Dutch-speaking firms. In doing so, this current study makes various substantial contributions to the IS literature. First, this study constructed and validated a comprehensive EA-based capability and treating it as a dynamic capability. Using the 16 measurement items across three dimensions (i.e., EA sensing, mobilizing, and transforming capability), this study helps researchers conduct more systematic analyses on the organization's EA-based capabilities. Second, this study empirically showed that the dynamic enterprise architecture capabilities construct is a crucial antecedent of business-IT alignment and process innovation. Furthermore, the latter two fully mediate the effect of dynamic enterprise architecture capabilities on organizational benefits. In doing so, this study expands upon previous EA-based capabilities and IT-enabled capabilities studies (Hazen et al., 2017; Rai & Tang, 2010; Shanks et al., 2018). Specifically, the identified mechanisms, and thus the mediating forces, through which benefits are achieved have theoretical relevance since a substantial amount of scholarship work under the assumption that the sole development of EA's with associated artifacts is a sufficient condition to enable business transformation and attain organizational benefits (Gong & Janssen, 2019; Kotusev, 2019). This study shows that organizational benefits as a result of EA-based capabilities can be achieved through intermediate-capabilities and IT-business benefits. These current findings might explain why firms still encounter organizational and externally imposed obstacles with realizing EA's intended business outcomes (Brosius et al., 2018; Löhe & Legner, 2014). The unfolded indirect effect of dynamic enterprise architecture capabilities on organizational benefits is also consistent with previous work on dynamic capabilities and their indirect effect on firm performance, see, for example, (Protogerou, Caloghirou, & Lioukas, 2012; Wilden et al., 2013).

This study provides business and IT managers with a potent source of value. First, firms should focus on dynamic enterprise architecture capabilities as an effective mechanism for promoting business-IT alignment and thus provide a better understanding of business processes and IS/IT, their interdependencies, and possible synergies. Dynamic enterprise architecture capabilities help cultivate the EA to reconfigure the business successfully and the IS/IT landscape, recombine resource and to adjust for and respond to unexpected changes an can thus be considered an essential driver for business-IT alignment. Another important managerial implication of this work is that firms can enable process innovation by deploying unique combinations of complementary resources and cooperation of cohesive units governed by dynamic enterprise architecture capabilities. Hence, dynamic enterprise architecture capabilities need to be positioned within the firm in such a way that they enable both alignment and process innovation, thereby using EA-based capabilities to their full potential. It is crucial that managers and decision-makers avoid the common fallacy of solely investing in the development of EA while not leveraging their EA-based capabilities to achieve enhanced levels of innovativeness and alignment, and ultimately high-levels of organizational benefits. Finally, the study findings suggest that, contrary to studies that used 'questionable' EA artifacts and framework, the dynamic enterprise architecture capabilities scale provides a reliable, valid, and useful diagnostic, (self)assessment and benchmarking tool grounded in theory.

### 6.2 Limitations and future work

Several study limitations guide future work, despite the attractiveness of the developed model and the assessments using reliable cross-sectional data. First, this study used self-reported data to test the hypothesized relationships in the research model. In doing so, it uses a similar approach as the studies of (Brosius et al., 2018; Hazen et al., 2017; Schmidt & Buxmann, 2011; Shanks et al., 2018) as objective measures are difficult to obtain. Although considerable time and effort were undertaken to account for possible measurement errors and bias, CMV may still be a concern as both the dependent and focal explanatory variables are perceptual measures derived from the same respondent (i.e., single informant). This study, currently, did also not triangulate the self-reported measures with, e.g., potentially available archival data from public sources. Including these additional data (e.g., financial measures) could help further validate the overall validity of the empirical outcomes as perceptual data are strongly correlated to objective measures (Shanks et al., 2018; Wu et al., 2015). This research encourages further research undertakings using current insights as a foundation. First, it would be valuable to look at the possible conditioning role of environmental turbulence, as previous studies have demonstrated its impact on organizational benefits (Pavlou & El Sawy, 2011; Rai & Tang, 2010; Tallon & Pinsonneault, 2011). Finally, this study concurs with Shanks et al. (Shanks et al., 2018) in that longitudinal research could lead to an enhanced understanding of dynamic enterprise architecture capabilities and the process of obtaining organizational benefits.